\documentclass[useAMS,usenatbib]{mn2e}
\usepackage{graphicx}

\title[Magnetar's non-thermal emission]{Radio emission of magnetars driven by the quasi-linear diffusion.}

\author[Osmanov Z.]{Osmanov Z.\thanks{E-mail:
z.osmanov@freeuni.edu.ge}\\
School of Physics, Free University of Tbilisi, 0159 Tbilisi,
Georgia}
\begin{document}


\pagerange{\pageref{firstpage}--\pageref{lastpage}} \pubyear{2009}

\maketitle

\label{firstpage}

\begin{abstract}
In this, paper we study the possibility of generation of
electromagnetic waves in the magnetospheres of radio magnetars by
means of the quasi-linear diffusion (QLD). Considering the
magnetosphere composed of the so-called beam and the plasma
components respectively, we argue that the frozen-in condition will
inevitably lead to the generation of the unstable cyclotron modes.
These modes, via the QLD, will in turn influence the particle
distribution function, leading to certain values of the pitch
angles, thus to an efficient synchrotron mechanism, producing radio
photons. We show that for three known radio magnetars the QLD might
be a realistic mechanism for producing photons in the radio band.

\end{abstract}


\section{Introduction}

Pulsars since their discovery deserve a great attention and many
aspects of their nature still remain unclear. One of the peculiar
class of the pulsars are characterized by long period of rotation,
which in turn leads to very strong magnetic fields exceeding the
so-called Schwinger limit, $B_{cr}\approx 4.41\times 10^{13}$G. For
such a huge magnetic field they are called magnetars. Despite some
predictions, that magnetars must be dark in the radio band
\citep{bhard}, it is now observationally evident that some of the
magnetars are detected in the radio band as well \citep{mcatalog}.
According to the McGill catalogue, there are six magnetars that
exhibit radio spectra. In particular, the following is the list of
the objects seen in the indicated radio band: 1E 1547.0-5408:
$[1.8-8.6; 18.5; 43; 45]$GHz, PSR J1622-4950: $[1.4–9.0; 17;
24]$GHz, and SGR J1745-2900: $[1.2–8.9; 14.6–20; 22]$GHz. There are
also three more magnetars, for which the radio emission has been
announced: 4U 0142+61: $0.11$GHz, XTE J1810-197: $[0.06; 0.35–19;
42; 88.5; 144]$GHz and 1E 2259+586: $[0.06; 0.11]$GHz \citep{malof},
although the detections have not yet been confirmed by another
observatory.

In this paper, we focus on the radio-loud pulsars to see the
possible role of the quasi-linear diffusion (QLD) in generation of
radio emission. For explaining the radiation in the radio band, we
account for the synchrotron emission process. On the other hand,
since in the magnetospheres of magnetars magnetic fields are very
strong, the corresponding energy loses are very efficient and for
studying the synchrotron radiation one has to take into account a
certain mechanism balancing the dissipative factors. Without
dissipation the pitch angles very soon become zero, leading to the
one dimensional distribution function of particles and as a result
the synchrotron mechanism completely vanishes. In this paper we rely
on the pulsar emission model developed by \citet{machus1,lomin}.
According to this approach, in the pulsar magnetospheres the
cyclotron instability appears \citep{kmm}, which during the
quasi-linear stage, causes a diffusion of particles along and across
the magnetic field lines, leading to the required balance.

The mechanism of QLD was applied to pulsars and active galactic
nuclei in a series of papers: \citep{malmach,o13,oc13,cmo13,co12}.
In the framework of the mechanism, the synchrotron radiation appears
by means of the feedback of the cyclotron waves on relativistic
particles due to the diffusion, and as a result, the pitch angles
are arranged according to the aforementioned balance. Generally
speaking, during the QLD, the physical system will be characterized
by two processes: (a) generation of the cyclotron waves and b) the
synchrotron mechanism.

In this paper we consider the magnetospheric parameters of
radio-loud magnetars to investigate the role of the QLD in
generation of the radio waves. Thus paper is organized as follows:
in Section 2 we introduce the mechanism of the QLD, in Sect. 3 we
apply the method to magnetars and obtain results, and in Sect. 4 we
summarize them.

\section{Main consideration} \label{sec:consid}

We assume that the pulsar's magnetosphere is composed of the
so-called primary beam with the Lorentz factor, $\gamma_b$ and the
bulk component with the Lorentz factor, $\gamma_p$
\citep{oc13,cmo13,co12}. By \citet{kmm} it was shown that in the
pulsar magnetospheric plasmas, which satisfy the frozen-in
condition, the anomalous Doppler effect induces resonance unstable
cyclotron waves
\begin{equation}\label{res}
\omega - k_{_{\|}}c-k_xu_x-\frac{\omega_B}{\gamma_{b}} = 0
\end{equation}
with the corresponding frequency \citep{malmach}
\begin{equation}\label{nu}
\nu \approx \frac{\omega_{_B}}{2\pi\delta\gamma_b},\;\;\;\;\; \delta
= \frac{\omega_p^2}{4\omega_{_B}^2\gamma_p^3},
\end{equation}
where $k_{_{\|}}$ is the longitudinal (along the magnetic field
lines) component of the wave vector, $u_x\approx
c^2\gamma_b/(\rho\omega_{_B})$ is the so-called curvature drift
velocity, $c$ is the speed of light, $\rho$ is the magnetic fields'
curvature radius, $k_x$ is the wave vector's component along the
drift, $\omega_{_B}\equiv eB/mc$ is the cyclotron frequency, $e$ and
$m$ are the electron's charge and the rest mass respectively,
$\omega_p \equiv \sqrt{4\pi n_pe^2/m}$ is the plasma frequency,
$n_p$ is the plasma number density and $B$ is the magnetic
induction. Inside the magnetosphere of magnetars the induction of
magnetic field on any lengthscale is given by
\begin{equation}\label{magn}
B\approx 3.2\times
10^{14}\times\left(\frac{P}{10s}\right)^{1/2}\times\left(\frac{\dot{P}}{10^{-11}ss^{-1}}\right)^{1/2}\times
\left(\frac{R_{st}}{R}\right)^3G,
\end{equation}
where $R_{st}\approx 10^6$cm is the magnetar's radius, $R$ is the
distance from the star's centre, $P$ is the rotation period and
$\dot{P}$ - the slow down rate of the magnetar. As it is clear from
equation (\ref{magn}), we have normalized $P$ and $\dot{P}$ on their
average values and as a result, the magnetic field becomes higher
than the Schwinger limit.  It is worth noting that the
aforementioned expression is derived from the standard assumption
that the spin-down rate is caused by the magneto-dipole emission.
Although based on observations of two anomalous pulsars
\cite{malofeev} suggest to revise the magnetar model itself.
According to this approach magnetars are young pulsars with
relatively low magnetic field.

In order to study the development of the QLD, one should note that
two major forces control dissipation. When particles emit in the
synchrotron regime, they undergo the radiative reaction force ${\bf
F}$, having the following components \citep{landau}:
\begin{equation}\label{fs}
    F_{\perp}=-\alpha_{s}\frac{p_{\perp}}{p_{\parallel}}\left(1+\frac{p_{\perp}^{2}}{m^{2}c^{2}}\right),
    F_{\parallel}=-\frac{\alpha_{s}}{m^{2}c^{2}}p_{\perp}^{2},
\end{equation}
where $\alpha_{s}=2e^{2}\omega_{_B}^{2}/3c^{2}$ and $p_{\perp}$ and
$p_{\parallel}$ are the transversal (perpendicular to the magnetic
field lines) and longitudinal (along the magnetic field lines)
components of the momentum respectively.

In nonuniform magnetic field, electrons also experience a force
${\bf G}$, that is responsible for conservation of the adiabatic
invariant, $I = 3cp_{\perp}^2/2eB$. The corresponding components of
${\bf G}$ are given by \cite{landau}:
\begin{equation}\label{g}
G_{\perp} = -\frac{cp_{\perp}}{\rho},\;\;\;\;\;G_{_{\|}} =
\frac{cp_{\perp}^2}{\rho p_{\parallel}},
\end{equation}
where $\rho$ is the curvature radius of the magnetic field lines.

The wave excitation leads to a redistribution process of the
particles via the QLD, which is described by the following kinetic
equation \citep{machus1}
\begin{eqnarray} \label{qld1}
    \frac{\partial\textit{f }}{\partial
    t}+\frac{1}{p_{\perp}}\frac{\partial}{\partial p_{\perp}}\left(p_{\perp}
    \left[F_{\perp}+G_{\perp}\right]\textit{f }\right)=\nonumber \\
    =\frac{1}{p_{\perp}}\frac{\partial}{\partial p_{\perp}}\left(p_{\perp}
D_{\perp,\perp}\frac{\partial\textit{f }}{\partial
p_{\perp}}\right),
\end{eqnarray}
where $\textit{f }$ is the distribution function of the zeroth
order, $D_{\perp,\perp}=D\delta |E_k|^2$, is the diffusion
coefficient, $|E_k|^2$, is the energy density per unit of wavelength
and $D=e^{2}/8c$ \citep{cmo10}. For estimating $|E_k|^2$, it is
natural to assume that half of the plasma energy density,
$mc^2n_b\gamma_b/2$ converts to the energy density of the waves
$|E_k|^2k$ \citep{machus1}, then for $|E_k|^2$ we obtain
\begin{equation}\label{ek2}
|E_k|^2 = \frac{mc^3n_b\gamma_b} {4\pi\nu},
\end{equation}
where $n_b$ is the number density of the beam.
%

It is evident from the model that it depends on relativistic effects
of electrons. Therefore, one has to estimate the maximum possible
values of particles' Lorentz factors. For this purpose we apply the
method of centrifugal acceleration \citep{osm7}, where, based on the
fact that plasma particles in strong magnetic field are in the
frozen-in condition, they follow magnetic field lines. On the other
hand, if these field lines are corotating, the particles will
undergo centrifugal force, efficiently amplified close to the light
cylinder surface (a hypothetical zone, where the linear velocity of
rotation exactly equals the speed of light), leading to extremely
efficient acceleration in the mentioned area. It has been shown by
\cite{rm00} that the Lorentz factors of centrifugally accelerated
electrons depend on the radial distance as $\gamma\approx
1\left(1-R^2/R_{lc}^2\right)^{-1/2}$, where $R_{lc}\equiv cP/(2\pi)$
is the light cylinder radius. As it is evident from this expression,
by reaching the light cylinder the energy must tend to infinity. In
real astrophysical situations, particles, apart from the centrifugal
force undergo other forces reducing the acceleration. In particular,
the electrons encounter soft photons and by means of the inverse
Compton (IC) scattering they lose energy. Initially energy losses
are negligible and acceleration is dominant, but in due course of
time energy gain and energy losses will balance each other and
further acceleration will be stopped. One can show that the maximum
Lorentz factor provided by the IC losses is given by \citep{osm7}
$$\gamma_{_{IC}}\approx 10^{14}\left[\frac{6\Omega}{U_{rad}(R_{lc})}\right]^2
\approx$$
\begin{equation}\label{gic}
\approx 1.05\times
10^{15}\times\left(\frac{P}{10s}\right)^2\times\left(\frac{10^{33}ergs\
s^{-1}}{L}\right)^2,
\end{equation}
where $\Omega=2\pi/P$ is neutron star's angular velocity of
rotation, $U_{rad} = L/(4\pi R_{lc}^2c)$ is the radiation energy
density on the light cylinder lengthscales and $L$ is the luminosity
of the magnetar.

Another mechanism, responsible for limiting the maximum attainable
energies is the breakdown of the bead on the wire (BBW)
approximation. In strong magnetic fields particles follow the field
lines, also gyrating around them. Apart from the Lorentz force, that
is responsible for gyration the electrons also undergo the force
perpendicular to the magnetic field lines, $F_{_{\perp}} =
m\Omega\left(2\gamma\frac{dR}{dt}+R\frac{d\gamma}{dt}\right)$.
Unlike this force, the Lorentz force, ${\bf F_{L}}=\frac{e}{c}{\bf
\upsilon_{rel}\times B}$, in due course of motion changes its
direction with respect to $F_{_{\perp}}$. Since the Lorentz force is
responsible for binding the particle close to the field lines, the
electrons follow them until $F_{_{\perp}}$ exceeds $F_L$. When this
happens, the bead-on-the-wire approximation is no longer valid and
acceleration vanishes, limiting the maximum attainable energies of
particles. As it has been shown in \citep{rm00} and generalized by
\cite{osm7} the corresponding maximum Lorentz factor is given by
$$\gamma_{_{BBW}}\approx \left[\frac{B_{lc}e}{2m\Omega
c}\right]^{2/3} \approx$$
\begin{equation}\label{gbbw}
\approx 1.19\times
10^{5}\times\left(\frac{P}{10s}\right)^{-3/2}\times\left(\frac{\dot{P}}{10^{-11}ss^{-1}}
\right)^{1/3}.
\end{equation}
We see from the aforementioned expressions that maximum Lorentz
factor governed by the IC mechanism is greater than that of the BBW
process. Therefore, the mechanism responsible for particle
acceleration in the magnetospheres of magnetars is the BBW and the
corresponding value of the Lorentz factor for the mentioned
parameters is of the order of $10^5$. This is the primary beam with
the Goldreich-Julian number density \citep{GJ}
\begin{equation}\label{nb}
n_{_{GJ}} = \frac{B} {Pce},
\end{equation}
 It is worth noting that
curvature emission also can limit the maximum energies of electrons,
but like the IC mechanism it is also small compared to the BBW.

Turning to equations (\ref{fs},\ref{g}) and by taking into account
the relations $\psi\equiv p_{\perp}/p_{\parallel}$,
$p_{\parallel}=mc\gamma_b$, one can estimate the following ration
\begin{equation}\label{FG}
\frac{F_{\perp}}{G_{\perp}}\approx 3.2\times
10^{-3}\psi^2\times\left(\frac{P}{10s}\right)^{-11/2}\times\left(\frac{\dot{P}}
{10^{-11}ss^{-1}}\right)^{4/3}.
\end{equation}
It is clear from equation (\ref{FG}) that for physically reasonable
parameters, one can neglect the transversal component of the
radiation reaction force. This regime differs from that of
considered in our previous study \citep{cmo13,cmo10}, because, as it
is evident from the above equation, the ratio is very sensitive to
the period of rotation and for long period neutron stars, unlike the
previously studied millisecond pulsars, the radiative force becomes
small compared to $G_{\perp}$.

Therefore, equation (\ref{qld1}) reduces to
\begin{eqnarray} \label{qld}
    \frac{\partial\textit{f }}{\partial
    t}+\frac{1}{p_{\perp}}\frac{\partial}{\partial p_{\perp}}\left(p_{\perp}
    G_{\perp}\textit{f }\right)=\nonumber \\
    =\frac{1}{p_{\perp}}\frac{\partial}{\partial p_{\perp}}\left(p_{\perp}
D_{\perp,\perp}\frac{\partial\textit{f }}{\partial
p_{\perp}}\right).
\end{eqnarray}

Equation (\ref{qld}) makes evident that two major factors compete in
this "game". On the one hand, the force responsible for conservation
of the adiabatic invariant attempts to decrease the transversal
momentum (thus the pitch angle), whereas the diffusion process, by
means of the feedback of the cyclotron waves, attempts to increase
the transversal momentum. Dynamically this process saturates when
the aforementioned factors balance each other. Therefore, it is
natural to study the stationary regime, $\partial\textit{f }
/\partial t=0$ and examine a saturated state of the distribution
function. After imposing the condition $\partial\textit{f }
/\partial t=0$ on Eq. (\ref{qld}) one can straightforwardly solve it
\begin{equation}\label{f}
    \textit{f}(p_{\perp})=C exp\left(\int \frac{G_{\perp}}{D_{\perp,\perp}}
    dp_{\perp}\right)=Ce^{-\left(\frac{p_{\perp}}{p_{\perp_{0}}}\right)^{2}},
\end{equation}
where $C={\it const}$ and
\begin{equation}\label{p0}
     p_{\perp_{0}}\equiv\left(\frac{2\rho D_{\perp,\perp}}{c}\right)^{1/2}.
\end{equation}
Since $\textit{f}$ is a continuous function of the transversal
momentum, it is natural to examine an average value of it and
estimate the corresponding mean value of the pitch angle,
$\bar{\psi}\equiv \bar{p}_{\perp_{0}}/p_{\parallel}$,
\begin{equation}\label{pitch}
\bar{\psi}
 = \frac{1}{p_{\parallel}}\frac{\int_{0}^{\infty}p_{\perp} \textit{f}(p_{\perp})dp_{\perp}}{\int_{0}^{\infty}\textit{f}(p_{\perp})dp_{\perp}}
\approx \frac{1}{\sqrt{\pi}}\frac{p_{\perp_{0}}}{p_{\parallel}}.
\end{equation}
As we see, the QLD leads to a certain distribution of particles with
the pitch angles, which will inevitably result in the synchrotron
radiation mechanism with the following frequency \citep{Lightman}
\begin{equation}
\label{nusyn} \nu_{syn}\approx 2.9\times
10^{-3}B\gamma_b^2\sin\bar{\psi}GHz.
\end{equation}
\begin{figure}
\resizebox{\hsize}{!}{\includegraphics{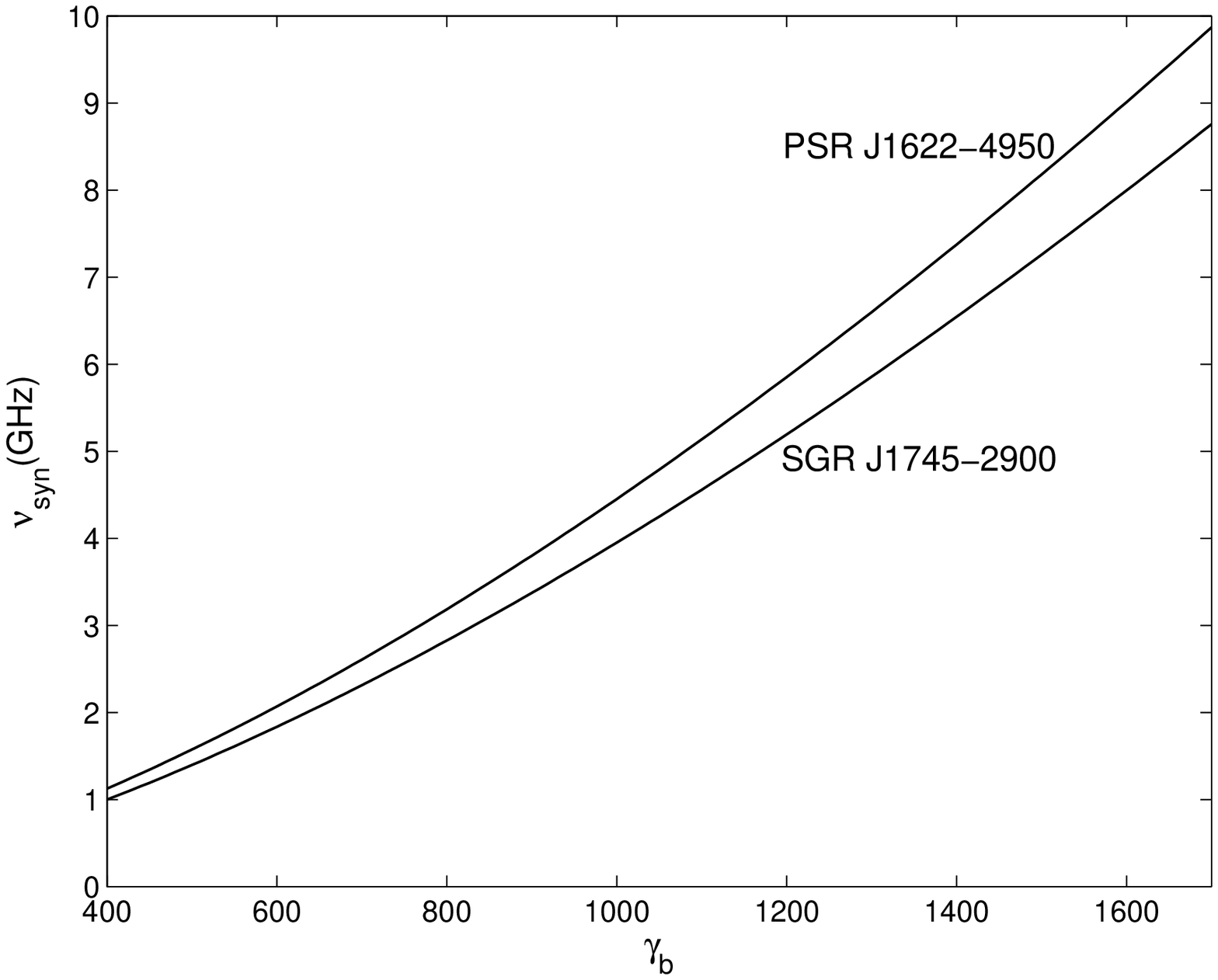}}
\resizebox{\hsize}{!}{\includegraphics{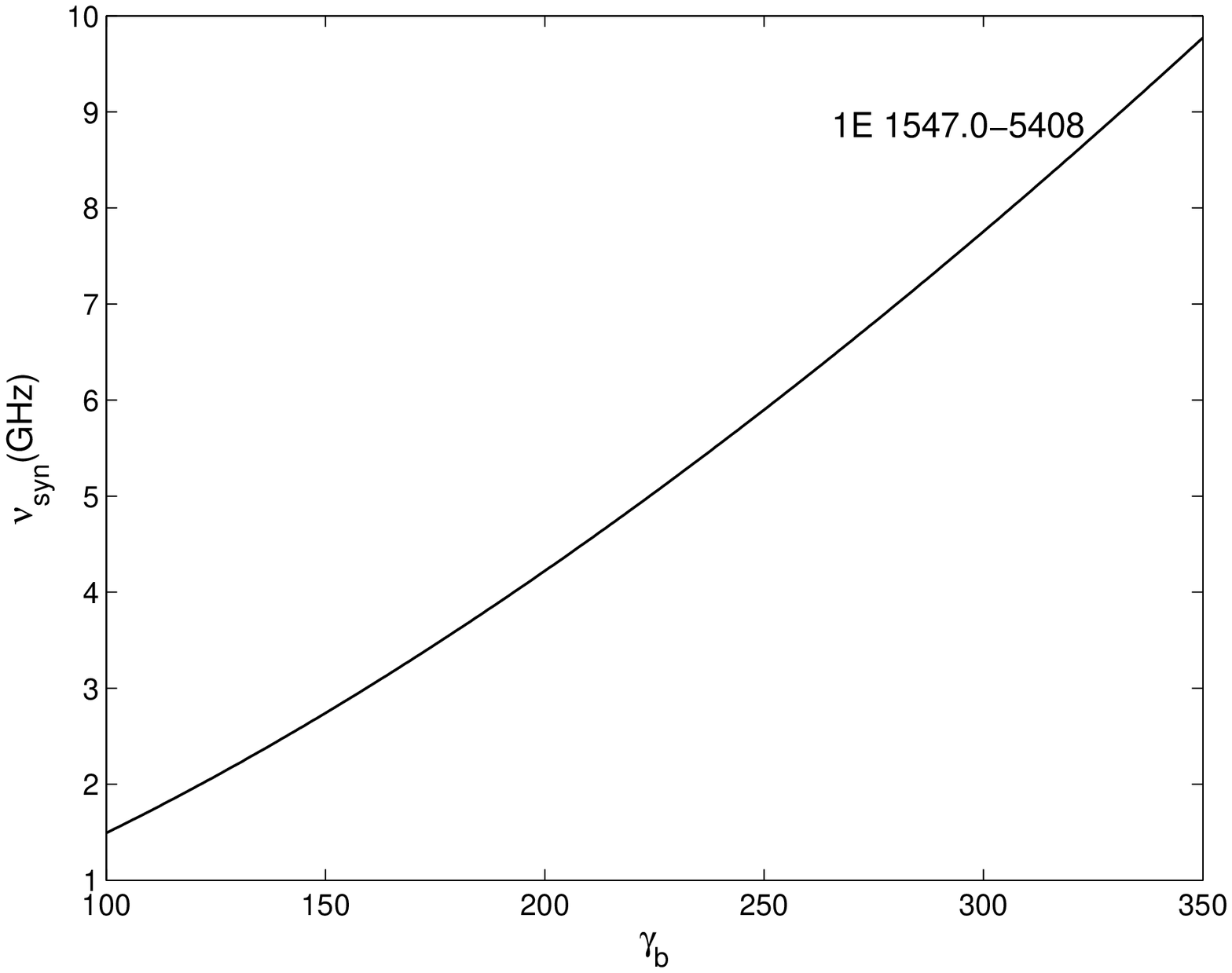}}
 \caption{Synchrotron photon frequency as a function of $\gamma_b$ for three
 radio magnetars: SGR J1745-2900; PSR J1622-4950 (top panel) and 1E 1547.0-5408 (bottom panel
 ). The
set of parameters is: $\gamma_p=2$, $R_{st}\approx 10^6$cm, $P=
3.76$s, $\dot{P} = 6.61\times 10^{-12}$ss$^{-1}$ for SGR J1745-2900;
$P= 4.33$s, $\dot{P} = 1.7\times 10^{-11}$ss$^{-1}$ for PSR
J1622-4950 and $P= 2.07$s, $\dot{P} = 4.77\times 10^{-11}$ss$^{-1}$
for 1E 1547.0-5408..} \label{fig1}
\end{figure}

\section{Discussion}
In this section we will apply the mechanism of the QLD to the long
period pulsars for studying the possibility of generation of radio
waves. In the framework of the proposed model, the QLD is provided
by the feedback of the unstable cyclotron waves. Therefore, it is
important to estimate the growth rate of the cyclotron instability.
\citet{kmm} have shown that for $\gamma_b/(2\rho\omega_B)\ll\delta$
(which is the case) the increment characterizing amplification of
the cyclotron waves is given by
\begin{equation}
\label{inc1} \Gamma=\frac{\omega_b^2}{2\nu\gamma_b},
\end{equation}
where $\omega_b \equiv \sqrt{4\pi n_be^2/m}$ is the plasma frequency
corresponding to the beam component. By considering mildly
relativistic particles of the plasma component with $\gamma_p=2$ and
the beam component with $\gamma_b = [10^2-10^5]$, by taking into
account that the energy is uniformly distributed,
$n_b\gamma_b\approx n_p\gamma_p\approx n_{_{GJ}}\gamma_{_{BBW}}$,
one can show that for the aforementioned parameters, the growth rate
is very high $\sim [10^{5}-10^{8}]$s$^{-1}$. Therefore, the
corresponding time-scale, $\tau\sim 1/\Gamma$, will be in the
following interval $\sim [10^{-8}-10^{-5}]$s. On the other hand, the
kinematic, or escape time-scale, $t_{esc}\sim R_{lc}/c$ is of the
order of $\sim 1$s. As we see, the instability time-scale exceeds by
many orders of magnitude the kinematic time-scale, which means that
the process is extremely efficient and therefore, physically
feasible.

As we have shown in the previous section, the cyclotron waves
inevitably influence the particle distribution via diffusion
(feedback mechanism) leading to certain pitch angles (see equation
(\ref{pitch})), which in turn leads to the synchrotron radiation. In
Fig. \ref{fig1} we show the dependence of synchrotron photon
frequency on the values of Lorentz factors for three radio
magnetars: SGR J1745-2900; PSR J1622-4950 and 1E 1547.0-5408. The
set of parameters is: $\gamma_p=2$, $R_{st}\approx 10^6$cm, $P=
3.76$s, $\dot{P} = 6.61\times 10^{-12}$ss$^{-1}$ for SGR J1745-2900;
$P= 4.33$s, $\dot{P} = 1.7\times 10^{-11}$ss$^{-1}$ for PSR
J1622-4950 and $P= 2.07$s, $\dot{P} = 4.77\times 10^{-11}$ss$^{-1}$
for 1E 1547.0-5408. It is clear from the plot that $\nu_{syn}$ is a
continuously increasing function of $\gamma_b$'s. This is a direct
consequence of equations (\ref{nu},\ref{ek2},\ref{p0}-\ref{nusyn}).
In particular, according to equation (\ref{nusyn}) the photon
frequency behaves as to be $\nu_{syn}\sim\gamma_b^2\overline{\psi}$,
on the other hand, by taking into account the relation
$D_{\perp,\perp}=D\delta |E_k|^2$, one can see from equations
(\ref{ek2},\ref{p0},\ref{pitch}) that $\overline{\psi}\sim
\gamma_b^{1/2}$, which by combining with equation (\ref{nusyn})
leads to the following dependence $\nu_{syn}\sim\gamma_b^{3/2}$.

As it is clear from the plots, for physically reasonable parameters,
the cyclotron instability can lead to generation of radio waves in
the observed interval of frequencies. We show only relatively low
range of frequencies: ($\sim [1-10]$GHz) and it is quite
straightforward to find physical parameters for higher frequencies
of radio emission. Therefore, the present investigation shows that
the contribution of QLD in generation of radio waves in radio
magnetars might be important.

\section{Summary}\label{sec:summary}

The main aspects of this work can be summarized as follows:
\begin{enumerate}

      \item In this paper we examined the role of the QLD
      in producing radio emission in the magnetospheres of
      three known radio magnetars: SGR J1745-2900; PSR J1622-4950 and 1E 1547.0-5408.

      \item Considering the anomalous Doppler effect, which leads to
      the unstable cyclotron waves, we have studied the feedback of
      these waves on a distribution of relativistic particles.
      Solving the equation governing the QLD, the corresponding
      expression of the average value of the pitch angle is derived
      and analysed for physically reasonable parameters.

      \item We have shown that for
      appropriate parameters $\gamma_p=2$, $\gamma_b=[3.5\times 10^2-1.7\times 10^3]$
      the QLD might
      provide a generation of radio emission in plasmas located on
      the light cylinder distances and might explain the observed frequencies
      in the interval $\sim [1-10]$GHz.

      \end{enumerate}

The present investigation shows that the QLD is a mechanism that can
explain a generation of radio waves in three confirmed radio
magnetars \citep{mcatalog}. The aim of this paper was to examine
only one part of the problem, although a complete study requires to
investigate the spectral pattern of emission as well. In the
standard theory of the synchrotron emission it is assumed that due
to the chaotic character of the magnetic field lines
\citep{bekefi,ginz}, the pitch angles lie in a broad interval (from
$0$ to $\pi/2$). In our model the distribution function of particles
is strongly influenced by the process of the QLD and as a result the
pitch angles are restricted by the balance of dissipative and
diffusive factors. This will inevitably lead to a spectral pattern,
different from that of \citet{bekefi,ginz}. Therefore, we will
investigate this problem in future studies.

In the framework of the model the synchrotron mechanism is
maintained by means of the induced cyclotron instability. On the
other hand, it is well known that in certain cases unstable
Cherenkov-drift modes might do the same work as well. In particular,
\cite{oc13}, examining the generation of synchrotron radiation by
means of the feedback of Cherenkov-drift modes on particle
distribution in magnetospheres of AGN, have shown high efficiency of
the process. Therefore, sooner or later we are going to study this
particular problem in the context of magnetars.

\section*{Acknowledgments} I thank professor G. Machabeli for valuable
discussions. This work was partially supported by the Shota
Rustaveli National Science Foundation grant (N31/49).

\bsp

\label{lastpage}

\end{document}